\documentstyle[12pt]{article}

\begin{document}

\begin{flushright}
gr-qc/0106004 \\
May 2001\\
\end{flushright}

\begin{centering}
\bigskip
{\leftskip=2in \rightskip=2in

{\large \bf $~~~~~~~$ Status of Relativity with observer-independent
$~~~~~~~$ length and velocity scales\footnote{Lecture 
given at the ``37th Karpacz Winter School of Theoretical Physics",
5-15 February 2001, Karpacz, Poland (to appear in the proceedings).}}}\\
\bigskip
\bigskip
\bigskip
{\bf Giovanni AMELINO-CAMELIA}\\
\bigskip
Dipartimento di Fisica, Universit\'{a} ``La Sapienza", P.le Moro 2,
I-00185 Roma, Italy\\ 
\end{centering}

\vspace{1cm}
\begin{center}
{\small \bf ABSTRACT}
\end{center}

{\leftskip=0.6in \rightskip=0.6in

I have recently shown that it is possible to formulate the Relativity 
postulates in a way that does not lead to inconsistencies in the case 
of space-times whose structure is 
governed by observer-independent scales of both velocity and length.
Here I give an update on the status of this proposal,
including a brief review of some very recent developments.
I also emphasize the role that one of the $\kappa$-Poincar\'{e} 
Hopf algebras could play in the realization of a particular
example of the new type of postulates. I show
that the new ideas on Relativity require us to extend
the set of tools provided by $\kappa$-Poincar\'{e}
and to revise our understanding of certain already available
tools, such as the energy-momentum coproduct.
%
%
%
%
}

\newpage
\baselineskip 12pt plus .5pt minus .5pt
\pagenumbering{arabic}
\pagestyle{plain} 

\section{Relativity and observer-independent scales}
In these notes I examine the status of my recent proposal~\cite{dsr1,dsr2} 
attempting to identify consistent Relativity postulates that involve
both an observer-independent velocity scale
($c \sim 3 {\cdot} 10^8 m/s$) and an observer-independent
length scale ($L_p \sim 1.6 {\cdot} 10^{-35}m$).
Readers already familiar with the proposal~\cite{dsr1,dsr2} 
might find anyway useful my review for what concerns
the results obtained in Refs.~\cite{jurekdsr,joaosteph,starkappa,gacJR}
which were motivated by Refs.~\cite{dsr1,dsr2}
and provided important contributions to the programme.

I start with a few remarks on the motivation for exploring
the possibility that the Relativity postulates might
involve an observer-independent length scale,
in addition to the now familiar
observer-independent velocity scale $c$.
The fact that the Planck length $L_p$ is proportional 
to both $\hbar$, the Planck constant, and $G$, the 
gravitational constant ($L_p \equiv \sqrt{\hbar G/c^3}$),
appears to invite one to speculate that $L_p$ might play a role
in the microscopic (possibly quantum) structure of space-time, and 
in fact many ``quantum-gravity" theories~\cite{rovhisto,gacqm100} 
have either assumed or stumbled upon this possibility.
However, a fundamental role for $L_p$ in the structure of space-time 
appears to be conceptually troublesome for one of the cornerstones
of Einstein's Special Relativity:
FitzGerald-Lorentz length contraction.
The Relativity Principle demands that physical laws should
be the same in all inertial frames, including the law
that would attribute to the Planck length a fundamental role
in the structure of space-time, whereas,
according to FitzGerald-Lorentz length contraction, 
different inertial observers would attribute different
values to the same physical length.
If the Planck length only has the role we presently attribute
to it, which is basically the role of a coupling constant
(an appropriately rescaled version of the coupling $G$),
no problem arises for FitzGerald-Lorentz contraction,
but if we try to promote $L_p$ to the status of an intrinsic
characteristic of space-time structure (or a characteristic of
the kinematic rules that govern particle propagation in space-time)
it is natural to find conflicts with FitzGerald-Lorentz contraction.

For example, it is very hard (perhaps even impossible)
to construct discretized versions or non-commutative versions
of Minkowski space-time which enjoy ordinary 
Lorentz symmetry.\footnote{Pedagogical illustrative examples of
this observation have been discussed, {\it e.g.}, 
in Ref.~\cite{hooftlorentz} for the case of discretization
and in Refs.~\cite{kpoinfirst,rueggnew,majrue,kpoinap} 
for the case of non-commutativity.}
Discretization length scales and/or non-commutativity length
scales naturally end up acquiring different values for
different inertial observers, just as one would expect 
in light of the mechanism of FitzGerald-Lorentz contraction.
Therefore, unless the Relativity postulates are modified,
it appears impossible to attribute to the Planck length
a truly fundamental (observer-independent) intrinsic
role in the microscopic structure of space-time.

We are of course not forced to introduce such a modification
of the Relativity postulates. In fact, we do not (yet?) have
any data that require us to attribute to $L_p$
an observer-independent role in the microscopic structure of space-time
(note, however, the intruiging indications emerging from
the data analysed in Ref.~\cite{gactp2} and references therein)
and the theoretical arguments suggesting such a role are still
rather debatable (see, however, the related comments reported here
in Section~4).
On the other hand, just because such an hypothesis is fully
consistent with presently available data (the presently accepted 
version of the Relativity postulates has been succesfully tested
only up to scales that are insufficient for probing the Planckian
regime) and because some, however preliminary, supporting theoretical
arguments have been found, it is of course legitimate to explore the
possibility that indeed the fundamentally correct formulation
of the Relativity postulates might involve another observer-independent
scale, in addition to $c$.

In order to prepare the ground for the line of analysis advocated 
in these notes (and previously advocated in Refs.~\cite{dsr1,dsr2})
it si convenient to review the role that observer-independent scales 
(or absence thereof)
already played in Galilean Relativity and Einstein's Special Relativity.
The Relativity Principle demands that ``the laws of physics are
the same in all inertial frames" and clearly 
the implications of this principle
for space-time structure and kinematics depend very strongly on whether 
there are
fundamental scales of velocity and/or length. In fact, the introduction
of a fundamental scale is itself a physical law, and therefore the
Relativity Principle allows the introduction of such fundamental
scales only if the rules that relate the observations performed
by different inertial observers are structured in such a way that
all inertial observers can agree on the value and physical interpretation
of the fundamental scales. 
The Galileo/Newton rules of transformation between 
inertial observers can be easily obtained by combining the 
Relativity Principle with the assumption 
that there are no observer-independent
scales for velocity or length.
For example, without an observer-independent velocity scale,
there is no plausible alternative~\cite{dsr1} to the simple
Galilean law $v'=v_0+v$ of composition of velocities.

Special Relativity describes the implications
of the Relativity Principle for the case in which there is
an observer-independent velocity scale. 
Einstein's second postulate can be naturally divided in two
parts: the introduction of an observer-independent
velocity scale $c$
and the proposal of a physical interpretation of $c$ as the
speed of light. This second postulate, when combined with the 
Relativity Principle 
(which is the first postulate of Special Relativity)
and with the additional assumption
that there is no observer-independent length scale
leads straightforwardly to the now familiar Lorentz transformations,
with their associated familiar formulation 
of FitzGerald-Lorentz contraction.
The assumption that there is no observer-independent length scale
plays a key role already in the way in which the second postulate
was stated. Experimental data available when Special Relativity was
formulated, such as the ones of the Michelson-Morley experiments,
only concerned light of very long wavelengths (extremely long
in comparison with the length scale $L_p$ introduced by Planck
a few years earlier) and therefore the second postulate could
have accordingly attributed to $c$ the physical role of
speed of long-wavelength light (the infinite-wavelength limit
of the speed of light); however, the implicit assumption of
absence of an observer-independent length scale allowed
to extrapolate from Michelson-Morley data a property
for light of all wavelengths. 
In fact, it is not possible to assign
a wavelength dependence to the speed of light
without introducing either a ``preferred" class of inertial frames 
or an observer-independent length scale.

All the revolutionary elements of Special Relativity
(in comparison with the Relativity of Galileo and Newton)
are easily understood as direct consequences of the
introduction of an observer-independent velocity scale.
This is particularly clear for the deformed law
of composition of 
velocities, $v'=(v_0+v)/(1+v_0v/c^2)$,
and the demise of absolute time (an absolute concept of time is 
untenable when an observer-independent velocity
scale governs the exchange of information between clocks).

Within the perspective here being adopted it is clear
that the Planck-length problem I am concerned with can be
described as the task of showing that the Relativity
Principle can coexist with some types of postulates stating that
the fundamental structure of space-time  
involves both an observer-independent velocity scale $c$
and an observer-independent length scale $L_p$.
The addition of an observer-independent length scale
does not require
major revisions of the physical interpretation of $c$, but, 
because of the mentioned connection 
between wavelength independence and absence of an observer-independent
length scale, I shall not authomatically assume that it is
legitimate to extrapolate from our long-wavelength data:
\begin{itemize}
\item{(law1):} The value of the fundamental velocity scale $c$
can be measured by each inertial observer as 
the $\lambda/L_p \rightarrow \infty$ limit of the speed of light
of wavelength $\lambda$.
\end{itemize}

While for $c$ we can at least rely on long-wavelength data,
we basically have no experimental information on the role (if any)
of $L_p$ in space-time structure. The type of exploratory
research programme I have proposed~\cite{dsr1,dsr2}
must therefore naturally start by identifying some examples of 
postulates that are logically consistent and involve both $c$
and $L_p$ as observer-independent scales.
Eventually one would like this programme to evolve to the point
where all such logically consistent formulations of Relativity
are identified, and then leave to experimentalists the final
task of establishing which (if any) of these candidates is realized
in Nature. Since this research programme is just starting off,
I felt~\cite{dsr1,dsr2} that it would be appropriate to focus on 
one specific illustrative example of new postulates, analysing it 
in depth so that we could be reassured that the set of such
logically consistent formulations of Relativity is non-empty.
This illustrative example of new Relativity postulates is introduced
in the next Section, and is also the main focus of most of the 
remainder of these notes.

\section{An illustrative example of the new type of $~~~~~~~~$
Relativity postulates}

As anticipated in the preceding Section, in this Section I will
focus on one example of new Relativity postulates.
For the framework I am advocating the only ingredient on which
we lack experimental guidance is the role to be attributed
in the postulates to $L_p$.
The other parts of the postulates have in fact already been
discussed in the preceding Section: the new theory will maintain
the Relativity Principle, it will introduce both $c$ and $L_p$
as observer independent scales in the postulates, and it will attribute 
to $c$ the physical role of the long-wavelength limit of
the speed of light.

In choosing an illustrative example of postulate attributing
a role to $L_p$ in space-time structure and kinematics,
I found~\cite{dsr1,dsr2} appropriate to give priority to ideas that
would have significant phenomenological consequences
(so that I could show explicitly that the issue I am considering
is not merely of academic interest)
and that can make some contact with preliminary indications
of quantum-gravity theories.
As I shall here discuss in greater detail 
in Sections~4 and 5
the hypothesis that the conventional dispersion relation $E^2= c^2 p^2$
be deformed at the Planck scale finds some motivation in recent
quantum-gravity theoretical studies and can lead to new effects
that are small enough to be consistent with all presently-available
data, while being large enough to be tested in the near future.
In the following I will also argue that, within the new type of Relativity 
theory which I am proposing, a deformation of the dispersion relation
can also be connected with the emergence of a minimum length,
another popular ``quantum-gravity idea".

Motivated by these considerations, in Refs.~\cite{dsr1,dsr2}
I chose to use the following 
illustrative example of postulate attributing
a role in space-time structure and kinematics to 
a length scale ${\tilde L}_p$:
\begin{itemize}
\item{(law2):} Each inertial observer can establish
the value of ${\tilde L}_p$ (same value for all inertial observers)
by determining the dispersion relation for photons, 
which takes the form $E^2 = c^2 p^2 + f(E,p;{\tilde L}_p)$, where 
the function $f$ has leading ${\tilde L}_p$ dependence given 
by: $f(E,p;L_p) \simeq {\tilde L}_p c E p^2$.
\end{itemize}
Of course, at the conceptual level we should even contemplate
the possibility that ${\tilde L}_p$ be completely unrelated
to $L_p$ (we cannot exclude the existence of a completely new
length scale in the correct formulation of the Relativity postulates),
but in light of the considerations reported above it appears
reasonable to explore in particular the possibility
the quantity\footnote{As illustrated
by the specific example (law2), the observer-independent
length scale must not necessarily have the physical meaning
of the length of something. For example, as indeed it happens in (law2),
the role of $L_p$ in space-time structure could be such that
it provides a sort of reference scale for momenta (wavelengths).} 
setting the strength of the dispersion-relation
deformation and the Planck length calculated {\it a la} Planck
be identified up to a numerical coefficient not too different from $1$
and a possible sign choice (${\tilde L}_p \equiv \rho L_p$,
with $\rho \in R$, $|\rho| \sim 1$).

I must now check the logical consistency of a Relativity theory
based on (law1) and (law2), and attempt to extract its most
characteristic features.

\subsection{Transformation rules (one-particle case)}

The logical consistency of the 
new postulates (law1) and (law2) requires
that, in their analyses of photon data 
in leading order in ${\tilde L}_p$, 
all inertial observers agree on
the dispersion relation $E^2 = c^2 p^2 + {\tilde L}_p c p^2 E$,
for fixed (observer-independent) values of $c$ and ${\tilde L}_p$.
The postulates do not explicitly concern massive particles,
which are at rest ($p=0$) in some inertial frames and in those frames
have a ``rest energy" which we indentify with the mass $c^2 m$.
For massive particles I tentatively adopt the
dispersion relation $E^2 = c^4 m^2 + c^2 p^2 + {\tilde L}_p c p^2 E$,
which satisfies these properties.
I postpone to future studies the possibility that these
postulates might coexist with more complicated dispersion
relations for massive particles of the 
type $E^2 = c^4 m^2 + c^2 p^2 + {\tilde L}_p c p^2 E + F(p,E;m;{\tilde L}_p)$,
which are consistent with (law2) not only in the case $F=0$
(here considered) but also whenever $F$ is such 
that $F(p,E;0;{\tilde L}_p)=F(p,E;m;0)=F(0,E;m;{\tilde L}_p)=0$.

Let me therefore assume $E^2 = c^4 m^2 + c^2 p^2 + {\tilde L}_p c p^2 E$
and look for boost generators (generators of rotations clearly do
not require modification) for which this dispersion relation
is an invariant (indeed valid for all inertial observers).
At this stage (see the wording adopted in (law2))
we shall be satisfied with checking logical consistency
at leading order in ${\tilde L}_p$.
For additional simplicity, here let me also 
limit\footnote{More general boosts can then be constructed 
by insisting~\cite{dsr1} on ordinary rotational invariance of the theory,
which still holds.}
my considerations to boosts along the $z$ direction of particles with 
momentum only in the $z$ direction, so that I am only to enforce
invariance of $E^2 = c^4 m^2 + c^2 p_z^2 + {\tilde L}_p c p_z^2 E$.
The Lorentz $z$-boost generator, 
$B_z = i c p_z {\partial / \partial E} 
+ i c^{-1} E {\partial / \partial p_z}$, 
clearly requires a deformation.
I make the ansatz 
$B_z^{{\tilde L}_p} = 
i [c p_z + {\tilde L}_p \Delta_E] {\partial / \partial E} + i 
[E/c + {\tilde L}_p \Delta_{p_z}] {\partial / \partial p_z}$,
for which one easily finds that the sought invariance translates
into the requirement 
$2 E \Delta_E - 2 p_z \Delta_{p_z} = -2 E^2 p_z - p_z^3$.
The simplest solutions are of the type 
$2 \Delta_E = 0~,~~~ \Delta_{p_z} = E^2 + p_z^2/2$
and 
$\Delta_{p_z} = 0 ~,~~~\Delta_E = - E p_z - p_z^3/(2 E)$.
Various arguments of simplicity~\cite{dsr1} (including considerations
involving combinations of boosts and rotations and the desire
to have generators which would be well-behaved even off shell)
lead me to adopt the first option, so the new $z$-boost generator
takes the form
\begin{equation}
B_z^{{\tilde L}_p} 
= i c p_z {\partial \over \partial E} + i 
[E/c - {\tilde L}_p E^2/c^2 
- {\tilde L}_p p_z^2/2] {\partial \over \partial p_z}
~.
\label{boosnew}
\end{equation}

One important observation to be made at this point is that the 
generators of boosts
(and rotations) constructed in the way I just described
turn out to correspond to 
the leading-order-in-${\tilde L}_p$
version of the Lorentz-sector generators of 
a well-known $\kappa$-Poincar\'{e} Hopf algebra~\cite{kpoinfirst,rueggnew,kpoinap}, 
the example of $\kappa$-Poincar\'{e} 
Hopf algebra first introduced in Ref.~\cite{majrue}.
In this sense just like the introduction of the Special Relativity
postulates led to preexisting Lorentz-group mathematics,
the example of new Relativity postulates I am analyzing
leads to preexisting $\kappa$-Poincar\'{e} mathematics
(note however that some of the observations reported 
in Subsection~2.3
do not fit in the $\kappa$-Poincar\'{e} mathematics, at least not
in the way in which it is presently understood).

For brevity, here I do not note the formulas for finite transformations.
Having obtained the new generators of boosts and rotations
one immediately obtains infinitesimal transformations
({\it e.g.}, ${dE / d\xi} = i [B_z^{{\tilde L}_p},E] = - c p_z ~,~~~
{dp_z / d\xi} = i [B_z^{{\tilde L}_p},p_z] = 
-E/c + {\tilde L}_p E^2/c^2 + {\tilde L}_p p_z^2/2$),
and then finite transformations are obtained
by straightforward (but tedious) integration.
The interested reader can find this discussion,
including explicit formulas for finite transformations, 
in Ref.~\cite{dsr1}.

\subsection{Length contraction}

The example of new postulates I am focusing on makes a non-trivial
assumption about energy-momentum space: even in the Planck
regime energy-momentum space is classical (although deformed).
This is a plausible, but strong, assumption, which of course
is reasonable to consider, especially in light of the exploratory 
attitude of these first studies of new Relativity postulates.
It might however be too much to assume that
also the space-time sector remains classical.
In this respect it is particularly important that in the 
preceding Subsection I was led to generators which had already
emerged in preexisting $\kappa$-Poincar\'{e} mathematics;
in fact, the relevant Hopf algebra has been understood~\cite{majrue,kpoinap}
as being dual to a non-commutative space-time, the
$\kappa$-Minkowski space-time ($l,m = 1,2,3$):
\begin{equation}
[x_m,t] = i {\tilde L}_p x_m ~,~~~~[x_m, x_l] = 0 ~.
\label{kmindef}
\end{equation}

This fact that the space-time counter-part of the energy-momentum
space which appears in the postulates might be ``quantum"
invites one to be prudent~\cite{dsr1} in making considerations
on the space-time picture of the transformation rules imposed
by the illustrative example of new Relativity postulates
on which I am focusing. We can however obtain some (partial)
information on the nature of this space-time sector even just using 
structures obtained in energy-momentum space. This is the task
that I reserved for the present Subsection.
My observations concern the general topic of ``relativistic
length contraction", considering both wavelengths (momenta)
and lengths.

One first observation concerns the 
possible emergence of a minimum wavelength 
(maximum momentum).
Let us consider a photon which, for a given inertial observer,
is moving along the positive direction of the $z$ axis with
momentum $p_0$ (and, of course, as imposed by the new dispersion
relation, has energy $E_0 \simeq p_0 + {\tilde L}_p c p_0^2/2$).
The new relativity postulates imply~\cite{dsr1} that 
for another inertial observer, which the first observer sees moving
along the same $z$ axis, the photon has momentum $p$
related to $p_0$ by
\begin{equation}
p = p_0 e^{-\xi} 
+ {\tilde L}_p p_0^2 e^{- \xi} 
- {\tilde L}_p p_0^2 e^{- 2 \xi} ~.
\label{finitep}
\end{equation}
Ordinary Lorentz boosts are of course
obtained as the ${\tilde L}_p \rightarrow 0$ 
limit of the new boosts (\ref{finitep}).
The comparison between (\ref{finitep})
and its ${\tilde L}_p \rightarrow 0$ limit provides 
some insight on the type of deformation of 
FitzGerald-Lorentz contraction that characterizes the new postulates.
As long as $p_0 < 1/|{\tilde L}_p|$ (wavelength $\lambda_0>|{\tilde L}_p|$)
and $e^{-\xi} \ll 1/(|{\tilde L}_p| p_0)$ the relation between $p$ and $p_0$
is well described by ordinary Lorentz transformations.
Within this analysis in leading order in ${\tilde L}_p$ 
it is not legitimate to consider the case $e^{-\xi} > 1/(|{\tilde L}_p| p_0)$
(which would require an exact all-order analysis of the implications of the
function $f(E,p;{\tilde L}_p)$ introduced in the postulates), 
but we can look at the behaviour
of the transformation rules when $e^{-\xi}$ is smaller but not
much smaller than $1/(|{\tilde L}_p| p_0)$.
While the transformation rules are basically unmodified 
when $e^{-\xi} \ll 1/(|{\tilde L}_p| p_0)$, as $e^{-\xi}$
approaches from below the value $1/(|{\tilde L}_p| p_0)$ 
the transformation rules are more and
more severely modified: for large boosts,
the ones that would lead to nearly Planckian wavelengths
in the ordinary special-relativistic case, the magnitude of
the wavelength contraction is significantly modified.
The modification takes the form of a reduction of the contraction
if ${\tilde L}_p > 0$.
For example, taking indeed ${\tilde L}_p >0$, 
for $e^{-\xi} \simeq 1/(3 {\tilde L}_p p_0)$ one
would ordinarily predict $p \simeq 1/(3 {\tilde L}_p)$ while the new
transformation rules predict 
the softer momentum $p \simeq 2/(9 {\tilde L}_p)$.
This suggests that there should exist an
exact all-order form of $f(E,p;{\tilde L}_p)$
(extending the present $f(E,p;{\tilde L}_p) \simeq \eta L_p c E p^2$ 
leading-order analysis) such that when one inertial observer assigns 
to the photon momentum smaller than $1/{\tilde L}_p$ 
(wavelength greater than ${\tilde L}_p$)
all other inertial observers also find momentum smaller 
than $1/{\tilde L}_p$.
It would then be possible to consider as unphysical momenta greater
than $1/{\tilde L}_p$. ${\tilde L}_p$ would have the role 
of observer-independent minimum wavelength.
This would not be very surprising for a Relativity with
observer-independent ${\tilde L}_p$ and $c$. ${\tilde L}_p$ would be the
observer-independent minimum wavelength, 
just in the same sense that $c$ is the observer-independent maximum
speed in ordinary Special Relativity (speeds greater than $c$
are unphysical and a velocity which is smaller than $c$ for one
inertial observer is also smaller than $c$
for all other inertial observers).

Similar findings emerge from the analysis of 
length contraction within the illustrative example of 
new Relativity theory on which I am focusing.
This can be shown by analysing
a gedanken length-measurement procedure.
A key point for this observation is the fact that the 
dispersion relation $E^2 \simeq c^2 p^2 + {\tilde L}_p c E p^2$ 
corresponds\footnote{The careful reader will realize that
by assuming that the relation $v = {dE / dp}$ is unmodified
I am actually stating a (perhaps not very strong)
property of the space-time sector. From the results obtained
in Ref.~\cite{gacmaj} one can conclude that this property is
enjoyed by the space-time of (\ref{kmindef}).}
to the deformed speed-of-light law
\begin{equation}
v_\gamma = {dE \over dp} = c \, (1 + {\tilde L}_p c^{-1} E)
~.
\label{eqvelocity}
\end{equation}
The wavelength dependence of the speed-of-light law (\ref{eqvelocity})
can lead to the emergence of a minimum length in measurement analysis.
In order to see this, let us consider two observers
each with its own (space-) ship moving in the same space direction, 
the $z$-axis, with different
velocities ({\it i.e.} with some relative velocity),
and let us mark ``A" and ``B" 
two $z$-axis points on one of the ships (the rest frame).
The procedure of measurement of the distance $AB$ is structured as 
a time-of-flight measurement:
an ideal mirror is placed at B
and the distance is measured
as the half of the time needed by a first photon wave packet,
centered at momentum $p_0$, sent from A toward B to be back at A
(after reflection by the mirror).
Timing is provided by a digital light-clock: another mirror is placed
in a point ``C" of the rest frame/ship, with the same $z$-axis coordinate
of A at some distance $AC$, and a second identical
wave packet,
again centered at $p_0$, is bounced back and forth between A and C.
The rest-frame observer will therefore measure $AB$ 
as $AB'=v_\gamma(p_0) {\cdot} N {\cdot} \tau_0/2$, where $N$ is the number of 
ticks done by
the digital light-clock during the A$\rightarrow$B$\rightarrow$A
journey of the first wave packet
and $\tau_0$ is the time interval corresponding to each
tick of the light-clock ($\tau_0 =2 \, AC/v_\gamma(p_0)$).
The observer on the second (space-) ship, moving with velocity $V$
with respect to the rest frame, will instead attribute to $AB$
the value
\begin{equation}
AB''= {v_\gamma(p)^2 - V^2 \over v_\gamma(p)} N {\tau \over 2}
~,
\label{dsfitzloa}
\end{equation}
where $p$ is related to $p_0$ through (\ref{finitep})
and $\tau$ is the  
time interval which the second observer, moving with respect to
the rest frame, attributes to each
tick of the light-clock.
It is easy to verify that $\tau$ is related to $\tau_0$ by 
\begin{equation}
\tau = {v_\gamma(p_0) \over \sqrt{v_\gamma(p')^2 - V^2}} \tau_0
~,
\label{tautau}
\end{equation}
where $p'$ is related to $p_0$ through the
formula for boosts in a direction orthogonal to
the one of motion of the photon.
Combining (\ref{dsfitzloa}) and (\ref{tautau}) one easily obtains
\begin{equation}
AB''= {[v_\gamma(p)^2 - V^2] v_\gamma(p_0) \over 
v_\gamma(p) \sqrt{v_\gamma(p')^2 - V^2}} 
N {\tau_0 \over 2} 
= {v_\gamma(p)^2 - V^2 \over 
v_\gamma(p) \sqrt{v_\gamma(p')^2 - V^2}}  AB'
~.
\label{dsfitzlob}
\end{equation}
The implications of (\ref{dsfitzlob}) for length contraction
are in general quite complicated, but they are easily analyzed in both
the small-$V$ and the large-$V$ limits
(examined here of course in leading order in $L_p$).
For small $V$ and small momentum (large wavelength) of the probes
Eq.~(\ref{dsfitzlob}) reproduces ordinary FitzGerald-Lorentz contraction.
For large $V$ Eq.~(\ref{dsfitzlob}) 
predicts that $AB''$ receives two most important
contributions: the familiar FitzGerald-Lorentz 
term ($AB' {\cdot} \sqrt{c^2-V^2}$) and a new term 
which is of order ${\tilde L}_p |p| AB'/ \sqrt{c^2-V^2}$.
As $V$ increases the ordinary FitzGerald-Lorentz contribution
to $AB''$ decreases as usual, but the magnitude of the new correction term
increases. Imposing\footnote{The careful reader will 
realize that actually one does not even need to impose $|p| > 1/AB''$
in order to find evidence of saturation of length contraction.
The new term of the form $|{\tilde L}_p| |p| AB'/ \sqrt{c^2-V^2}$
(for ${\tilde L}_p>0$) increases with increasing $V$ for two reasons:
because the denominator $\sqrt{c^2-V^2}$ gets smaller and because
the numerator gets larger (larger $V$ means larger boost and
therefore larger $p$). One easily then finds that at some point
(for some value of $V$) the correction term is actually larger
than the 0th-order term $\sqrt{c^2-V^2}AB'$. This is the point
where my leading-order analysis stops to be reliable, but clearly
the result suggests that length-contraction is saturating
to a minimum length.} 
$|p| > |\delta p| > 1/AB''$ (the probe wavelength
must of course be shorter than the distance being measured)
one arrives at the 
result $AB'' > \sqrt{c^2-V^2}AB' + {\tilde L}_p AB'/(AB'' \sqrt{c^2-V^2})$.
For ${\tilde L}_p$ positive this result clearly 
implies that $AB'' > {\tilde L}_p$ for all values of $V$.
Again I must remind the reader that I am working
in leading order in ${\tilde L}_p$, and therefore
the results cannot be trusted when $V$ is large enough that the
correction term is actually bigger
than the 0-th order contribution to $AB''$,
but we can trust the indications of this analysis as long
as the correction is smaller than the 0-th order term, and 
in that regime one finds that (for positive ${\tilde L}_p$)
FitsGerald-Lorentz contraction
is being significantly softened in the region corresponding 
to nearly Planckian contraction.
This result clearly supports the hypothesis that there should
exist a consistent all-order form of $f(E,p;{\tilde L}_p)$ 
such that when one 
inertial observer assigns to a length value greater than ${\tilde L}_p$ 
all other inertial observers also find that length to be greater
than ${\tilde L}_p$.
Such a form of $f(E,p;{\tilde L}_p)$ would provide a relativistic
theory with observer-independent scales $c$ and ${\tilde L}_p$ 
in which ${\tilde L}_p$
has the intuitive role of ``minimum length" described above.

\subsection{Kinematical conditions for particle-production processes}

As I emphasized in Ref.~\cite{dsr1}, an important requirement for
the logical consistency of a Relativity theory is
that the laws imposed on particle-production processes should be 
the same in all inertial frames,
{\it i.e.} all observers should agree on whether or not
a certain particle-production process is allowed.
This requirement is trivially satisfied in ordinary Special Relativity.
Let me discuss this in the simple case 
of a scattering process $a+b \rightarrow c+d$ (collision
processes with incoming particles $a$ and $b$ and outgoing
particles $c$ and $d$). 
Also in this Subsection for simplicity I focus on
the case of one space dimension (the generalization to
multi-dimensional spaces is described in Ref.~\cite{dsr1})
and work to leading order in ${\tilde L}_p$.
The special-relativistic kinematic requirements for 
such processes are $E_a + E_b - E_c -E_d=0$ and $p_a + p_b - p_c -p_d=0$,
and, using the special-relativistic transformation rules, $d E_j/d\xi = - p_j$,
$dp_j/d\xi = -E_j$, one immediately verifies that when the requirements
are satisfied in one inertial frame they are also verified in all
other inertial frames~\cite{dsr1}.

The fact that the energy-momentum transformation rules imposed
by the postulates (law1),(law2) are non-linear (unlike
special-relativistic transformation rules) provides room for
various alternatives for the laws to be satisfied
by particle-production processes. This point is discussed in
detail in Ref.~\cite{dsr1}. Here I just want to mention
two possibilities whose consistency with the postulates
has been already verified.
The first example is 
\begin{itemize}
\item{(cons):} $a+b \rightarrow c+d$ collision
processes must satisfy the requirements
\begin{equation}
E_a + E_b - {\tilde L}_p c p_a p_b
- E_c -E_d + {\tilde L}_p c p_c p_d = 0
~,
\label{conservnewe}
\end{equation}
\begin{equation}
p_a + p_b - {\tilde L}_p (E_a p_b + E_b p_a)/c 
- p_c -p_d + {\tilde L}_p (E_c p_d + E_d p_c)/c = 0
~.
\label{conservnewp}
\end{equation}
\end{itemize}
It is easy to verify, using (\ref{boosnew}), that these conditions
are satisfied in all inertial frames if they are satisfied in 
one of them.
The particle-production law (cons) is a rather natural manifestation
of the observer-independent scale ${\tilde L}_p$. The difference
between the special-relativistic particle-production laws
and the laws (cons) reflects the introduction of  ${\tilde L}_p$
in the postulates, just like the difference between
the Galilean velocity-composition law, $v'=v_0+v$,
and the special-relativistic velocity-composition 
law, $v'=(v_0+v)/(1+v_0v/c^2)$, reflects the introduction of $c$
in the postulates.

The type of non-linearity encoded in
the postulates (law1),(law2) is also consistent with a completely
different alternative type of particle-production laws.
Whereas (cons) is a ``single-channel conservation law",
just like its counter-part in Special Relativity, it is also
possible to find consistent multi-channel laws for particle production.
A significant example is
\begin{itemize}
\item{(cons'):} $a+b \rightarrow c+d$ collision
processes must satisfy one of the requirements 
obtained by permutations of $p_a,p_b,{\tilde p}_c,{\tilde p}_d$
in the conditions
\begin{equation}
E_a + E_b - E_c -E_d = 0
~,
\label{conservnewepr}
\end{equation}
\begin{equation}
p_a \dot{+} p_b \dot{+} {\tilde p}_c \dot{+} {\tilde p}_d =0
~,
\label{conservnewppr}
\end{equation}
where the deformed sum $\dot{+}$ is defined by
$k \dot{+} q \equiv k + q + {\tilde L}_p E_k q $
and ${\tilde k} \equiv - k - {\tilde L}_p E_k k$
(with $E_k$ denoting the energy that corresponds to the momentum $k$).
\end{itemize}
It is again easy to verify, using (\ref{boosnew}), that the
particle-production law (cons')
is satisfied in all inertial frames if it is satisfied in 
one of them. It is somewhat more difficult to build some intuition
for this alternative possibility which can be introduced 
consistently with the illustrative example of new Relativity postulates
on which I am focusing. However, if the expectation that the space-time
sector is described by (\ref{kmindef}) is correct, the law (cons')
is actually to be favoured. In fact, in Ref.~\cite{starkappa}
it was shown that the construction of a consistent theory
on the space-time (\ref{kmindef}) leads to the law (cons').

The content of (cons') is not as shocking as it may seem at first sight.
It says that, in the case of a $a+b \rightarrow c+d$ process,
there are 24 channels available to the process 
(associated with the 24 possible permutations of the 
particle momenta $p_a,p_b,{\tilde p}_c,{\tilde p}_d$). 
The process will be allowed whenever one of the 24 cases is satisfied. 
What can be somewhat shocking is that this type of structure does not allow
to describe the particle-production laws as laws of conservation
of energy-momentum (since the particles can choose between 24
different conditions, none of these conditions can acquire the
special status of a, deformed, law of energy-momentum conservation).
Reassuringly in the limit in which the particles have energies (and momenta)
much smaller than $1/|{\tilde L}_p|$ (the only limit in which
we presently have conclusive experimental information on energy-momentum
conservation) all 24 channels collapse into a single energy-momentum
conservation condition, the one of Special Relativity.

\subsection{Postulates beyond leading order}

The results described up to this point are the ones
on which I based\footnote{However,
the specific formulation (cons') of the second type
of particle-production rules is a more recent result
obtained in Ref.~\cite{starkappa}.}
my proposal~\cite{dsr1,dsr2} of Relativity postulates with
more than one observer-independent scale.
The analysis was done in leading order in ${\tilde L}_p$,
leaving for future studies the search of consistent
choices of the all-order 
function $f(E,p;{\tilde L}_p)$
that appears in the postulate (law2).
As mentioned in Subsection~2.2
(and in Refs.~\cite{dsr1,dsr2})
an important hint appears to come from the fact that
my analyses led to leading-order expressions for the
generators of boosts and rotations that are recognizable
as the leading-order approximation of the generators
in the Lorentz sector of the example of $\kappa$-Poincar\'{e} 
Hopf algebra proposed in Ref.~\cite{majrue}.
The connection between the new Relativity postulates and this
Hopf algebra was further explored, within an all-order analysis,
by Kowalski-Glikman~\cite{jurekdsr} who adopted as
a natural candidate for the function $f(E,p;{\tilde L}_p)$
the one which is inferred from the all-order form of
the relevant $\kappa$-Poincar\'{e} ``$M^2$" casimir
({\it i.e.} in analogy with my leading-order proposal).
Specifically, using the form of this casimir,
Kowalski-Glikman provided
an all-order generalization\footnote{Note that in generalizing to all orders
the minimum-length proposal I had put forward in Ref.\cite{dsr2},
Kowalski-Glikman also observed~\cite{jurekdsr} that upon describing my
function $f(E,p;{\tilde L}_p)$ with the $\kappa$-Poincar\'{e} ``$M^2$" casimir
one is led (upon making again the sign choice required for the emergence
of a minimum wavelength) to the conclusion that in the infinite-energy limit
the speed of massless particles is actually infinite. In my leading-order
analysis one could only reliably establish that in the range where the
leading order is meaningful the  speed of massless particles grows
with energy. Kowalski-Glikman's observation that this speed actually
diverges asymptotically is fully consistent with the conceptual
starting points of my work: if Planckian lengths should not be
subject to FitzGerald-Lorentz contraction there must be a particular
limit of the theory in which the transformation rules are effectively
Galilean (no contraction of Planckian lengths). It is however also 
important to stress (this was not stressed in Ref.~\cite{jurekdsr})
that the infinite-speed limit is only to be understood as an
asymptotic behaviour: any real photon (finite energy) will have
a finite speed and transformations between observers using such 
photons as their probes would not be Galilean. There is no role for
real Galilean transformations in the new theory, but the 
Galilean asymptote is crucial for the overall logical consistency.}
of one of the arguments (here reviewed
in Subsection~2.2) which I used~\cite{dsr2} in support of the emergence
of a minimum wavelength (maximum momentum) in the
illustrative example of new Relativity postulates considered
here and in Refs.~\cite{dsr1,dsr2}.
Bruno, Kowalski-Glikman and I also showed, in a very recent 
study~\cite{gacJR},
that the use of the relevant $\kappa$-Poincar\'{e} ``$M^2$" casimir
in the postulates leads to consistent transformation rules 
between different inertial observers, again generalizing to
all orders the leading-order results I reported in Refs.~\cite{dsr1,dsr2}.
An all-order formulation of the 
particle-production rules (cons') were reported by Arzano and myself
in the very recent Ref.~\cite{starkappa}.

All these results appear to support my conjecture that the results 
reported in leading order in Refs.~\cite{dsr1,dsr2} could be
straightforwardly generalized to the level of an all-order analysis.

\section{Relation with quantum symmetries}

It is natural to assume that the coexistence of the Relativity Principle
with length and velocity observer-independent scales should lead to
the emergence of space-time symmetries, just in the same sense that
Special Relativity, with its single observer-independent scale,
leads to Lorentz symmetry. The fact that the new symmetries should
involve an additional scale and should reproduce ordinary Lorentz invariance
in a certain limit of the additional scale (the $L_p \rightarrow 0$ limit)
suggests that the subject of ``quantum groups" and ``quantum algebras"
should be in some way relevant. 
It is probably too early to conclude that this connection should
characterize all examples of the type of relativistic theories
I proposed (theories in which the Relativity Principle coexists 
with observer-independent scales of both velocity and length), 
but it definitely characterizes the specific illustrative example I 
analyzed in detail here and in Refs.~\cite{dsr1,dsr2}.
In fact, as already observed in Section~2,
the postulate (law2) involves a dispersion
relation which corresponds to the leading-order-in-${\tilde L}_p$
version of a casimir that has emerged~\cite{majrue,kpoinap} 
in the quantum-algebra literature, and, upon imposing consistency
with the Relativity Principle, I was led to boost (and rotation)
generators which can also be recognized as
the leading-order-in-${\tilde L}_p$
version of the generators of the relevant quantum algebra.
Like the Special-Relativity postulates provided a possible role
in physics for the pre-existing mathematics of the Lorentz group, 
the illustrative example of Relativity with two
observer-independent scales I considered led me to a possible
role in physics for the $\kappa$-Poincar\'{e} quantum algebra proposed 
in Refs.~\cite{majrue,kpoinap}.

I was unable to find in the mathematics literature 
the finite new-boost trasformations I obtained in Ref.~\cite{dsr1},
but, based on comparison with the analysis in Ref.~\cite{rueggnew}
(which concerned a rather similar quantum algebra), 
I am confident that my results
have been derived consistently with
the spirit of quantum algebras.

While there is a wide-spread belief (see, {\it e.g.}, Ref.~\cite{rueggnew})
that $\kappa$-Poincar\'{e} quantum algebras do not have an associated
group action (this action should only lead
to a ``quasi-group" in the sense of Batalin~\cite{batalin}),
I have shown in Ref.~\cite{dsr1} that the Lorentz sector of the specific 
$\kappa$-Poincar\'{e} algebra~\cite{majrue,kpoinap} that appears to be
relevant for my illustrative example of new Relativity postulates
does reassuringly lead to ordinary group structure.
This follows straightforwardly from the observation 
that the Lorentz sector of the relevant $\kappa$-Poincar\'{e} algebra
does satisfy the criteria derived by Batalin~\cite{batalin}
for the associated finite transformations to form group.

While the one-particle sector appears to be fully consistent with
the mathematics of quantum algebras, the analysis of particle-production
processes reported in Section~2 appears to require some new 
algebraic tools.
In particular, at least according to the standard interpretation
of the strictly mathematical language of analysis of quantum algebras,
the mathematics literature would support 
the expectation~\cite{majrue,kpoinap}
that the composition of momenta in the two-particle sector
should involve a troubling lack of symmetry between pairs of particles.
Even in the case of two identical particles it appears necessary to 
handle to the two momenta in a nonsymmetric way, while the composition
of energies is undeformed.
On the contrary, the analysis reported in Section~2
shows that consistency with the postulates does not require
any such loss of symmetry under exchange of particles.
It appears therefore plausible that a mathematical description
of the line of analysis advocated in Section~2
may require the introduction of new concepts in the subject
of quantum algebras.

The very recent analysis reported in Ref.~\cite{starkappa}
appears to provide the tools for introducing these new concepts
in $\kappa$-Poincar\'{e}, and provide solutions for some of the
reasons of concern which have traditionally obstructed attempts
to apply the $\kappa$-Poincar\'{e} formalism in physics. To these observations
I devote the remainder of this Section.

\subsection{Role of the $\kappa$-Poincar\'{e} coproduct in 
particle-production rules}

One of the key obstacles for physics applications of the
$\kappa$-Poincar\'{e} formalism is associated with the $\kappa$-Poincar\'{e} 
coproduct, which readers familiar with $\kappa$-Poincar\'{e} will recognize
(of course, in leading order) in the ``$\dot{+}$" operation
here introduced in Subsection 2.3: two momenta, $p$ and $k$,
are combined in the coproduct by the rule $p \dot{+} k$.
The fact that $p \dot{+} k$ is affected by a severe loss
of $p,k$-exchange symmetry has motivated some skepticism
toward the applicability of $\kappa$-Poincar\'{e} in physics.

For example, in the $\kappa$-Poincar\'{e} literature it has been
assumed that scattering processes involving two incoming and
two outgoing particles should conserve total momentum in the sense
that the coproduct sum of the incoming momenta should equal 
the coproduct sum of the outgoing momenta. This appears troubling
since the lack of symmetry of the coproduct would imply, for example, 
that in the case of two identical particles colliding to produce two
other identical particles one should choose which of the incoming
momenta enters the coproduct from the left and a similar choice would
have to be made for the outgoing particles.
This problem is solved by the particle-production rule (cons')
here introduced in Subsection 2.3, which involves the coproduct in a way
that however does not force us to choose the ordering of the incoming
and outgoing momenta: (cons') treats in a fully symmetric way
all momenta involved in the process. The price payed for this reassuring
result is that (cons') does not admit interpretation as an ordinary
rule of energy-momentum conservation: (cons') actually states that, {\it e.g.},
a process with two incoming and two outgoing particles, rather than having
to obey a single fixed conservation rule, can be realized through
any one of 24 conservation rules, obtained by permutations of the
four momenta involved in the process.

This prediction is rather strikingly new, but it does not pose any
conceptual problem for application in physics (no required choice
between identical particles) and is actually consistent with
all available data, which concern momenta that are
much smaller than the inverse of the Planck length (in which case
the mentioned 24 particle-production channels all collapse into 
a single and ordinary conservation rule).

As a first example of application of the rule (cons') 
let me note here the explicit
formulas that according to (cons') would describe the process
in which a photon of energy $E$ and a photon of energy $\epsilon$,
with $\epsilon \ll E$, collide and produce an electron-positron pair.
The analysis of (cons') is rather simple 
if we assume that the process is at threshold (incoming photons
only barely satisfy the energetic requirements for producing
the electron-positron pair) and we include only the leading-order
corrections, of order ${\tilde L}_p E^2$. In this limit one easily
finds that the 24 channels actually all give raise to the same
conservation rule (again, I stress that this is the leading-order result). 
While in conventional physics one would impose
the relation $2 p = E-\epsilon$ on the common momentum of the
produced pair (at threshold the electron and the positron necessarily
emerge with identical momenta), according to (cons') 
one should impose the condition $2 p = E-\epsilon + {\tilde L}_p E^2/4$
(all 24 channels predict this same relation in leading order).
Combining this result with the structure of the dispersion relation
one finds that there is no leading-order
deformation of the threshold condition: 
the deformation of the dispersion relation is compensated
by the deformation of momentum conservation, giving
back the ordinary threshold condition $E \epsilon = m_e^2$,
with $m_e$ the electron mass.

This cancellation of leading-order corrections to the threshold
condition for processes described in a highly boosted frame
(a frame which is highly boosted with respect to the center-of-mass
frame) may at first appear reassuring. The new Relativity theory
(and its underlying $\kappa$-Poincar\'{e} mathematics) turn out to be
a deformation of conventional Relativity that is even milder
than expected (one could expect effects to be small because of the
Planck-length suppression, but one might have not guessed that
even the leading-order term in the Planck length cancels out).
However, it is instead more correct to describe this
result of cancellation of leading-order effect as disappointing.
In fact, just for processes seen in a ``LAB frame" which is highly
boosted with respect to the center-of-mass frame there is growing
evidence in support of an anomaly in the threshold conditions.
This evidence emerges from astrophysical observations which
I will discuss in Section~5, but the key point is that in order
to explain these observations it would have been useful~\cite{gactp2,gactp1}
to encounter a leading-order correction to the threshold conditions.

The next question to ask is of course: Does the cancellation of 
leading-order corrections to the threshold conditions mean that the
new Relativity theory cannot provide an explanation for these puzzling
observations? A definite answer to this question still requires additional
investigations. The new
Relativity theory might still explain the mentioned puzzling observations,
but, if it does, the structure of the solution must be more complicated
than a simple deformation of the threshold conditions.
It seems to me that there are at least three promising avenues to seek
a solution of the observational paradoxes within the new Relativity
theory: (i) The peak of the cross section for the particle-physics
processes~\cite{gactp2,gactp1} relevant for the mentioned observational
paradoxes is not exactly at threshold; it is somewhat above threshold.
The fact that the leading-order corrections cancel each other out
for the very special conditions required by threshold production
does not necessarily imply that above threshold one should find
a similar cancellation. This should be studied and compared
with the observations. (ii) The structure
of the new particle-production kinematical requirements of the new
Relativity theory may actually combine in non-trivial way when
a given process actually involves more than one microscopic
process ({\it e.g.} a small cascade). Again, this may affect
the comparison of the new theory with observations. (iii) As I shall
emphasize in the next Subsection, it appears that the new Relativity
theory requires careful handling of composite particles (particles
composed by a few fundamental particles). The mentioned astrophysical
observations are believed to involve~\cite{gactp2,gactp1} 
photons, electrons, protons and pions. 
In a Relativity theory applicable all the way
down to the Planck length it appears even conceivable that photons
and electrons (and quarks) might not be fundamental particles,
and certainly protons and pions cannot be treated as truly
fundamental particles. The observations I report in the next Subsection
suggest that, in the new Relativity theory, collisions involving
composite particles might behave quite differently from collisions
among fundamental particles. This is another ingredient
which should be taken into account in seeking an explanation
for the mentioned puzzling observations in astrophysics.

\subsection{Differences between microscopic and macroscopic bodies}

In the preceding Subsection I presented a solution for one
of the obstacles, concerning particle-production processes,
for physics applications of the
$\kappa$-Poincar\'{e} formalism. The solution was motivated by
my attempt of constructing new Relativity theories and the role
that $\kappa$-Poincar\'{e} might have in some of these new theories.
There is another, perhaps even more serious,
obstacle for physics applications of the
$\kappa$-Poincar\'{e} formalism: while deformed dispersion relations 
of the type $E^2 = c^2 p^2 +  {\tilde L}_p c E p^2$
are consistent with all available data on fundamental particles
(these data concern particles with energy-momentum which is
much smaller than $1/{\tilde L}_p$)
such a deformation is clearly unacceptable for macroscopic bodies
(a macroscopic body easily has energy-momentum that is
much greater than $1/{\tilde L}_p$ and our data on macroscopic
bodies are clearly inconsistent with the deformed dispersion relation).
If a deformed dispersion relation
of the type $E^2 = c^2 p^2 +  {\tilde L}_p c E p^2$
should play a role in physics, clearly its applicability must somehow
be confined to systems of one or a few fundamental particles;
it should not hold for macroscopic bodies.

In the way in which the $\kappa$-Poincar\'{e} formalism has been developed
until now there is no room for such a separation between microscopic
and macroscopic realms. This is again associated with the role that
the coproduct had been assumed to play in previous $\kappa$-Poincar\'{e} studies.
In fact, it was assumed that the coproduct should characterize the total
momentum of a multi-particle system; for example, a system
composed of two particles, one with momentum $p$ and the other with
momentum $k$, would be characterized by total momentum $p \dot{+} k$
(or $k \dot{+} p$, an alarming choice must be made also in this case)
and this is found to lead to total momentum and total energy
which transform just like the single-particle energy momentum, {\it i.e.}
following the deformed dispersion relation.

Also in this respect the proposal of (cons')
can be used to motivate a solution of the paradox, again inspired
by the idea that some of the new Relativity theories of the 
general type proposed in Ref.~\cite{dsr1,dsr2} might in some way
involve the $\kappa$-Poincar\'{e} formalism. The point is that the concept
of total momentum of a multi-particle system must, as all concepts
in physics, be introduced to reflect an operatively-defined 
property of a physical system. A natural opportunity for attributing
a physically meaningful significance to the concept of total
momentum is provided by collision processes: we will be able to give
operative meaning to the concept of total momentum of two incoming
particles if some combination
of the energy-momenta of these incoming particles is conserved 
in the process (the corresponding combination of the outgoing-particles 
energy-momenta takes the same value).
The rules (cons'), which are fully consistent with the new
Relativity postulates, do not admit this type of interpretation:
they cannot be described as an equality between a sum involving
only the incoming momenta and a sum involving
only the outgoing momenta. According to (cons') it is in particular
not legitimate to take the coproduct sum of the incoming momenta
as the total momentum of that two-particle system.
This is sufficient to provide the needed opportunity 
for a separation between microscopic
and macroscopic realms: if the total momentum is not identified
with the coproduct sum of the momenta it will not necessarily
obey the same dispersion relation of the energy-momentum of a single particle.
To a system composed of a number $N$ of particles in the new Relativity
theory we cannot assign a meaningful ``total momentum"; we must keep track 
of all individual momenta of the composing particles and analyze collision
processes from that starting point.

Let me also observe, however, that in some weak sense it is possible
to introduce some sort
of total momentum.
Using again (cons') it appears that we should describe the concept
of total momentum only as some sort of average property of
a macroscopic body. We will have a good definition of total momentum
of a macroscopic body if we identify a characteristic momentum
of a macroscopic body which is conserved in collisions between
macroscopic bodies. (cons') does not allow to enforce
such a condition exactly, but it does allow to introduce such a
condition in an appropriate statistical sense.
Let us consider the collision of two macroscopic bodies, each composed
by an Avogadro number of fundamental particles: such a collision
at a fundamental level will actually involve a very large number
of collisions between the fundamental particles that compose
the macroscopic bodies. Each of these microscopic collisions
will actually be characterized by a single one of the 24 channels
(when the process is $2 \rightarrow 2$) but the collision between
two macroscopic bodies will involve such a large number of these
microscopic collisions that it will be characterized by the
average of the 24 channels.\footnote{I am of course describing the
ingredients of this proposal in a simplified way: in a real situation
different types of fundamental particles would compose the macroscopic
body and different types of collisions would occur at a fundamental
level. However, the main point will still hold; the deformation described
by the rules (cons') would be averaged out as a result of the large
number of collisions occurring at the fundamental level.}

At least in leading order in ${\tilde L}_p$, it appears plausible
that this authomatic averaging procedure would lead to the
introduction of a (non-fundamental) concept of total momentum
of a macroscopic body which, just as in conventional physics,
is based on the ordinary sum of momenta.
This however might only be applicable to collisions between
macroscopic bodies whose velocities are not very high, so that a small
boost is sufficient to take the system to the center-of-mass frame.

In support for this possibility let me analyze
the implications of (cons') for a center-of-mass collision of 
two identical particles with momenta $p$ and $-p$
respectively, and of course same energy $E$, that produces two other
identical particles just above threshold. 
This microscopic process would be one of the many microscopic processes
that occur when two macroscopic bodies collide.
Conventional
physics would predict that the sum of the momenta of the outgoing particles,
$p_1' + p_2'$, should vanish: $p_1' + p_2' =0$.
From (cons') one easily finds
that, in leading order\footnote{In the previous Subsection I applied
(cons') to a microscopic process seen by an observer characterized
by a large boost with respect to the center-of-mass frame. In that case
the 24 channels agreed in leading order. As shown in the present Subsection,
in the description of the same microscopic process by the center-of-mass 
observer the 24 channels that characterize (cons') do not agree even
in leading order. It is
easy to check that this difference between center-of-mass observers
and highly boosted observers
is fully consistent with (and actually reflects the properties of)
the deformed boost action.} in this case the 24 channels that 
characterize (cons') split up into 8 channels with $p_1' + p_2' =0$,
4 channels 
with  $p_1' + p_2' + {\tilde L}_p (E/c) (p_1' - p_2')/2 = {\tilde L}_p E p/c$,
4 channels 
with  $p_1' + p_2' + {\tilde L}_p (E/c) (p_1' - p_2')/2 = - {\tilde L}_p E p/c$,
4 channels 
with  $p_1' + p_2' - {\tilde L}_p (E/c) (p_1' - p_2')/2 = {\tilde L}_p E p/c$,
4 channels 
with  $p_1' + p_2' - {\tilde L}_p (E/c) (p_1' - p_2')/2 = - {\tilde L}_p E p/c$.
A single process of this type would follow a single one of these 24 options,
but a collection of a large number of these processes would be primarily
characterized by the average behaviour of the 24 channels,
which is simply $<p_1' + p_2'> =0$.

\section{Relations with other results of quantum-gravity research}

Perhaps the most important implication of the proposal I put forward
in Refs.~\cite{dsr1,dsr2} is that it is now clear that there are
two alternatives for introducing the Planck length in the fundamental
structure of space-time.
Before the proposal~\cite{dsr1,dsr2} the only option was provided
by the present interpretation of the Planck length as a scale
characteristic of the rules of dynamics, just a rescaled value
of the gravitational coupling $G$. In that approach 
the Planck length could enter space-time structure 
only when accompanied by an associated background ({\it e.g.} as
a scale present in a/the vacuum solution of the equations
of dynamics). Indeed, even maintaining 
the postulates of Special Relativity unmodified (including 
Fitgerald-Lorentz length contraction), it is of course 
possible~\cite{dsr1,grbgac,gampul,emnnew}
that the Planck length be associated with some sort of background.
This would be analogous to the well-known 
special-relativistic description of the motion of an electron
in a background electromagnetic field, which
is described by different 
observers in a way that is consistent
with the Relativity Principle, but only when these
observers take into account
the fact that the background electromagnetic field
also takes different values in different inertial frames.
The Planck length could play a similar role in space-time structure,
{\it i.e.} it could reflect the properties of a background,
but then the presence of such a background
would allow to single out a ``preferred" class of inertial frames
for the description of the short-distance structure of space-time
(the ``preferred" class of inertial frames would of course be
identified using Fitgerald-Lorentz contraction which does not allow
the presence of an observer-independent scale in space-time structure).

The results I reported in Refs.~\cite{dsr1,dsr2} show that in addition
to this traditional scenario, which introduces the Planck
length together with a preferred class of inertial frames, 
it is also possible to follow another scenario
for the introduction of the Planck length. 
This second option does not predict
preferred inertial observers but does require 
a short-distance deformation of boosts and an associated
modification of the Relativity postulates.
My intuition that such a scenario should be explored 
found additional encouragement even after
the announcement of Refs.~\cite{dsr1,dsr2},
especially through conversations
in which I became aware of arguments put forward by
other colleagues~\cite{garay,sussboost} in support of the hypothesis
that we might eventually encounter a
deformation of boosts (of course, also those arguments were 
motivated~\cite{garay,sussboost}
by quantum-gravity issues, such as minimum length and the quantum
mechanics of black holes).

I must also stress that, in light of my results~\cite{dsr1,dsr2}, 
it appears necessary for authors to be
more careful in their description of certain popular quantum-gravity
concepts, such a ``minimum length" and ``deformed dispersion relation".
In many  quantum-gravity
approaches~\cite{mead,padma,da,ngmpla,gacmpla,vene,grossme,amaven,kab}
one or another formulation of the concept of ``minimum length"
is discussed. However, these studies do not clarify how the 
presence of a minimum length could affect boosts. This appears
to be a serious omission, since, as emphasized above, 
there are two options for
introducing such concepts: either as a characteristic of 
quantum geometrodynamics
(without any modification of the Special-Relativity postulates)
or as a characteristic of the Relativity postulates.
Similar issues arise in the analysis of 
approaches (see, {\it e.g.}, Refs.~\cite{grbgac,gampul})
predicting new-physics
effects that would be strong
for particles of wavelength of the order of the Planck length
but would be weak for particles of larger wavelengths,
such as the ones associated with deformed dispersion relations.
Clearly, assuming ordinary special-relativistic rules of transformation
of energy and momentum,
these dispersion relations would allow to select a preferred
class of inertial frames, but I have shown that deformed dispersion
relations can also be introduced as observer-independent laws,
at the price of revising Special Relativity.

Another class of studies which have emerged in more or less direct
connection with quantum-gravity research and might be reanalyzed
from the perspective advocated in my proposal~\cite{dsr1,dsr2}
is the one of deformations of various types of algebras
motivated by the desire to implement the existence of concepts
such as minimum length, minimum de Broglie wavelength 
or a maximum 
accelleration~\cite{maggiore,kempmang,dadebro,maxA1,maxA2}.
Again in these studies until now much emphasis has been placed
on the algebraic tools, but the readers were left without any explicit
remarks concerning the faith of the Special-Relativity postulates.
It would be interesting to reanalyse the relevant
proposals within the new relativistic conceptual framework here proposed,
particularly working toward the identification of transformation
rules such that the equations describing
minimum length and/or minimum de Broglie wavelength and/or 
maximum accelleration
acquire the status of being observer-independent (valid in every
inertial frame). What are then the new relativistic transformation
rules between observers? Are they physically acceptable?
(For example, do the new Lorentz transformations form group,
or just a quasigroup?)

\section{Closing remarks}
This closing Section is devoted to a summary of the main results
discussed in the previous Sections and to the discussion
of some developments of this new research line which,
in my opinion, deserve urgent attention.

\subsection{Relativity can be doubly special}

From the viewpoint advocated here and in Refs.\cite{dsr1,dsr2}
the Relativity Principle
is somewhat hostile to the introduction of observer-idependent
physical scales. In that respect, Einstein's Relativity postulates
well deserve to be qualified as ``special", since they provide an example
in which the Relativity Principle coexists with an 
observer-independent (velocity) scale.
My studies have shown that one can also consistently construct
a ``Doubly Special Relativity",
in which the Relativity Principle coexists with 
observer-independent scales of both length (or momentum) and velocity.
If this option, now shown to be viable, turns out to be chosen 
by Nature we would have a sort of ``Quantum Special Relativity",
in the sense encoded in the role played by the Planck length.

\subsection{Toward a new approach to Quantum Gravity}

While the motivation for my studies comes from the desire to
eventually unify General Relativity and Quantum Mechanics,
the approach is at present still only able to handle
flat space-time. In a sense I have constructed 
a ``Quantum Special Relativity" but the natural ultimate goal
of this research programme should be a ``Quantum General Relativity".

In working toward this ultimate objective a useful intermediate
step could be the one of applying the new postulates in
contexts with a curved, but still fixed (non-dynamical), space-time,
such as De Sitter or Schwarzschild.

Another interesting possibility is the one of describing space-time
curvature as a requirement of non-commuting momenta\footnote{An important
role in my interest in this possibility was played by exciting 
discussions with Shahn Majid at the time of our collaboration
for the study~\cite{gacmaj}.}. If this viewpoint turned out
to be correct, one could perhaps reach the formulation
of a ``Quantum General Relativity" by an appropriate extension
of the $\kappa$-Minkowski space-time (\ref{kmindef})
to some sort of ``$\kappa$" phase space (which however here is intended
as the space $x_i,t,p_i,E$ rather than just $x_i,p_i$).

Even before these preliminary steps are done, there are certain
conceptual issues that must be analyzed.
One key point is that at present the Planck length $L_p$ is
seen as a quantity which is derived from three fundamental constants
$c$, $G$ and $\hbar$. The fundamental constants $c$, $G$ and $\hbar$
already have their own operative definitions ($c$ can be measured
as the speed of long-wavelength photons,
$G$ can be measured, {\it e.g.},
by simple studies of the solar system, 
and $\hbar$ can be measured by combined analysis of data on
quantum effects, {\it e.g.}, the black-body spectrum and
the photoelectric effect).
If $L_p$ is introduced in the Relativity postulates then $L_p$
is authomatically promoted from the status of derived constant
to the status of fundamental constant (since it would then have
its own operative definition, as in (law2)).
The relation between $L_p$, $c$, $G$ and $\hbar$ would accordingly acquire
the very rare status of a relation between fundamental concepts,
all with their own operative definition. This is very rare
in physics, but it cannot be excluded since we have at least
one example in which something like this happens: the inertial mass
and the gravitational mass of a particle are two concepts with independent
operative definitions, but there is a relation between them (they are
equal to each other because of the Equivalence Principle).
Therefore one way to conceptualize my proposal~\cite{dsr1,dsr2}
would require a non-trivial step, somewhat analogous to the
introduction of the Equivalence Principle. 
There is of course another (perhaps
even more radical) conceptual alternative: somehow this formalism
that provides an intrinsic operative definition for the Planck length 
might eventually lead to the understanding of one of the two scales
not explicitly present in the new Relativity postulates, either $G$ 
or $\hbar$, as a derived concept. This is a fascinating possibility,
which however might require surprising discoveries in the
development of the formalism.

\subsection{Phenomenology}

It is important to notice that the possibility of new Relativity postulates
involving the Planck length is not merely of academic interest.
On the contrary, as shown by the illustrative example of new postulates
on which I focused, there can be significant phenomenological
implications.
These implications have been discussed in some detail in Ref.~\cite{dsr1}.

The deformation of the dispersion relation introduced in the 
postulate (law2) can be tested~\cite{grbgac,schaef,billetal,glast}
with forthcoming experiments, even if the deformation scale ${\tilde L}_p$ is
indeed of the order of the tiny Planck length ${\tilde L}_p \sim L_p$.

The deformed rules for particle production (here discussed in Subsection~2.3)
could be most effectively tested in experiments sensitive to the
structure of the threshold requirements for particle production.
Interestingly, some of these experiments,
observations of ultra-high-energy cosmic rays~\cite{refCR}
and of Markarian501 photons~\cite{refMK}, have recently obtained data
that appear to be in conflict with conventional theories and
appear to require~\cite{colgla,kifu,ita,aus,gactp2,gactp1}
a deformation of the kinematic conservation rules applied to
collision processes.
These observations clearly provide some encouragement for
the idea of new Relativity postulates. As I emphasized in Subsection~3.1,
the illustrative example of new postulates on which I focused
appears to provide an avenue for explaining these paradoxical
observations, but more work is needed in order to substantiate
this hypothesis. At a preliminary level of analysis I found
(in Ref.~\cite{dsr1} and here in Subsection~3.1) a result which 
does not provide clear encouragement for this hypothesis:
I found two corrections to the threshold conditions, and the corrections
do have exactly the right magnitude to explain the observational paradoxes,
but the two corrections cancel each other out (to the relevant order). 
In Ref.~\cite{dsr1} and here in Subsection~3.1 I identified certain
alternative mechanisms for explaining the paradoxes within the new
Relativity theory. A detailed analysis of these alternative mechanisms
is postponed to future studies. This is perhaps
the most exciting and urgent 
issue for the development of the new Relativity theory.

A third class of phenomenological studies that could be significantly
affected by the new Relativity postulates is the one pertaining
to cosmology and the early stages of evolution of the Universe.
The interested reader can find brief remarks on this point in
Ref.~\cite{dsr2} and a more detailed (and preliminarily quantitative) 
study in Ref.~\cite{joaosteph}.

\subsection{Other forms of new Relativity postulates}

My proposal~\cite{dsr1,dsr2} of exploring the possibility
of new Relativity postulates, involving the Planck length,
could of course be followed investigating a large
variety of classes of new postulates; however, until
now all results have been obtained within the one
illustrative example on which I focused also in this paper.

It would be interesting to explore a few alternative
possibilities. If nothing else, alternative choices of
the postulates could clarify whether or not the satisfactory
outcome of the consistency tests of the illustrative example 
considered in these first studies is significant.
If it turned out that other choices of this new type of postulates
lead to some inconsistencies, the illustrative example which
has proven to have such nice properties could be seen
as a strong candidate (while at present it must be only
considered as a first example of consistent new postulates).
In this respect a key point might emerge from the analysis
of combinations of boosts. Whereas in the illustrative example
pursued until now the new Lorentz transformations form group
in the ordinary sense, it appears plausible that
other choices of the new postulates would only lead
to quasigroup structure~\cite{batalin}, a rather undesireable
feature.

Another potentially interesting possibility is the one
of attempting to introduce even a third 
observer-independent scale in the postulates.
Since my results showed that a logically consistent framework can
emerge from Relativity postulates with a second observer-independent
scale, it is now natural to wonder whether a third 
observer-independent scale could also be consistently introduced. 
Motivated by the studies I reported in Refs.~\cite{dsr1,dsr2},
Kowalski-Glikman has 
briefly presented in Ref.~\cite{jurekdsr} some ``aestethic arguments"
(not guided by experimental input or by conceptual urgency, but
by an intuition for the conceptual elegance of the fundamental
laws of physics) in favour of Relativity with three 
observer-independent scales,
but did not formulate any attempt to provide an operative
definition of the third scale (the entire analysis reported in
Ref.~\cite{jurekdsr} relies on the type of deformation of the 
dispersion relation which I had introduced with (law2)).
Consistently with the explorative spirit of my proposal~\cite{dsr1,dsr2},
I neither favour nor disfavour {\it a priori} any particular number
of observer-independent scales. We know experimentally
that there is at last one
observer-independent scale, $c$, in the Relativity postulates.
My studies have shown that a second observer-independent scale 
can be consistently introduced, but we now must wait for the virdict
of experimental tests. Examples of Relativity postulates with three
observer-independent scales should be studied, and, if any class
of such postulates turned out to be logically consistent,
corresponding experimental tests are certainly well motivated.
In this respect I should emphasize that, as discussed in Ref.~\cite{dsr1},
for each observer-independent scale 
the postulates should also provide an operative definition
of that scale (this key point was omitted in Ref.~\cite{jurekdsr}).
Also important are some considerations related with the remarks
I made above in Subsection~5.2. In promoting the Planck length
from the status of derived scale to the status of fundamental
scale (with its own independent operative definition) we are faced
with significant (but exciting)
conceptual challenges, and of
course attempts of using my proposal~\cite{dsr1,dsr2} in the
direction of introducing even a third observer-independent
scale are confronted by 
conceptual challenges which are even more serious
({\it e.g.}, if the three scales in the postulates are related with $c$ 
and two combinations of the other scales $G$ and $\hbar$, we could 
even contemplate the possibility of 
interpreting both $G$ and $\hbar$ as derived scales!).

\subsection{Understanding $\kappa$-Poincar\'{e}}

The illustrative example of new Relativity postulates on which
I focused ended up making strong contact with pre-existing 
mathematics of  $\kappa$-Poincar\'{e} algebras, first developed through pioneering
studies of Lukierski, Ruegg and collaborators~\cite{kpoinfirst,rueggnew}
(although the relevant example of $\kappa$-Poincar\'{e} algebra was
discovered more recently~\cite{majrue,kpoinap}).
While waiting for experimental tests, 
one can only hope that somehow this connection with 
pre-existing mathematics might be a good auspice for the fortunes
of this type of new Relativity
postulates, just like the introduction, nearly a century ago,
of the Special-Relativity postulates, which turned out to
lead to pre-existing Lorentz mathematics, was blessed by a long
string of experimental successes.

While pre-existing Lorents mathematics really provided all the tools
needed for analyses based on Special Relativity, 
my proposal was confronted~\cite{dsr1,dsr2} with some missing pieces
in the development of $\kappa$-Poincar\'{e}, particularly the lack of
understanding of the role of the coproduct in the laws for
particle production and the lack of the needed mechanism 
for confining the applicability of the $\kappa$-Poincar\'{e}
dispersion relation to the microscopic realm.
The new Relativity postulates led me to propose some solutions
for these outstanding problems of $\kappa$-Poincar\'{e}.
In Subsection 3.1, using results obtained in Refs.~\cite{dsr1,starkappa},
I argued that the are some consistent ways to introduce 
the $\kappa$-Poincar\'{e} coproduct in particle-production rules,
without any of the feared problems associated with a lack of symmetry
under exchange of the momenta of identical particles.
In Subsection 3.2, using again results obtained in Ref.~\cite{starkappa},
I argued that the concept of ``total momentum of a multi-particle body"
is a highly non-trivial concept in $\kappa$-Poincar\'{e},
and that this allows to find ways to confine 
the applicability of the $\kappa$-Poincar\'{e}
dispersion relation to the microscopic realm.

These results, besides playing a key role in my new type of Relativity
theories, appear to have even wider significance, possibly of use
in all contexts in which $\kappa$-Poincar\'{e} is being considered 
as a useful mathematical structure.
The two problems solved in Subsections 3.1 and 3.2 were the most
alarming residual problems in the \underline{conceptual} 
analysis of $\kappa$-Poincar\'{e}.
Their solution should perhaps reenergize research aimed at addressing
the residual \underline{technical} challenges of $\kappa$-Poincar\'{e},
particularly the identification of a natural measure for integration
over energy-momentum space~\cite{gacmaj,starkappa}.

\section*{Acknowledgements}
I am indebted to several colleagues, starting of course
with my collaborators Michele Arzano, Rossano Bruno, 
Jerzy Kowalski-Glikman and Tsvi Piran.
I am also greatful to Jerzy Lukierski for conversations on
the subject of ``quasigroups", as described by Batalin~\cite{batalin},
and to Daniele Fargion and Maurizio Lusignoli for conversations on some 
of the particle-physics processes relevant for the paradoxical threshold
anomalies suggested by certain astrophysical observations.
Less technical (but still very useful from the point of view of
finding encouragement for my efforts)
conversations with Dhram Ahluwalia, Stephon Alexander, Maurizio Gasperini,
Joao Magueijo and Lee Smolin are also very greatfully acknowledged.
Finally, I should thank all the colleagues who expressed interest
in the proposal I put forward in Refs.~\cite{dsr1,dsr2}
and pointed to my attention some other formalisms and ideas which,
like the $\kappa$-Poincar\'{e} formalism, may turn out to provide
opportunities for the introduction of Relativity theories
of the type I proposed; in particular, I must thank
Dharam Ahluwalia for bringing to my attention the studies
in Refs.~\cite{kempmang,dadebro},
Roberto Aloisio for bringing to my attention some remarks
in Ref.~\cite{garay},
Michele Arzano for bringing to my attention some remarks
in Ref.~\cite{maggiore},
Jerzy Kowalski-Glikman for bringing to my attention Ref.~\cite{sussboost}
and Giuseppe Marmo for bringing to my attention Refs.~\cite{maxA1,maxA2}.

\baselineskip 12pt plus .5pt minus .5pt

\vfil

\end{document}